\documentclass[sigconf]{acmart}
\usepackage{algorithm}
\usepackage[noend]{algpseudocode}

\AtBeginDocument{%
  \providecommand\BibTeX{{%
    \normalfont B\kern-0.5em{\scshape i\kern-0.25em b}\kern-0.8em\TeX}}}

\copyrightyear{2019}
\acmYear{2019}
\setcopyright{acmlicensed}
\acmConference[EuroMPI'19]{EuroMPI'19: 25th European MPI User\'s Group Meeting}{10 - 13 September 2019}{ETH Z\"urich, Switzerland}
\acmBooktitle{EuroMPI'19: 25th European MPI User's Group Meeting, 10 - 13 September 2019, ETH Z\"urich, Switzerland}
\acmPrice{00.00}
\acmDOI{00.0000/0000000.0000000}
\acmISBN{000-0-0000-0000-0/00/00}

\begin{document}

%
\title[Strong scaling using deep pipelined Conjugate Gradients]{Improving strong scaling of the Conjugate Gradient method for solving large linear systems using global reduction pipelining}

%
\author{Siegfried Cools}
\authornote{University of Antwerp, Department of Mathematics and Computer Science, Building G, Middelheimlaan 1, BE-2020 Antwerp, Belgium.}
\email{siegfried.cools@uantwerp.be}
\affiliation{%
  \institution{University of Antwerp}
  \streetaddress{Middelheimlaan 1}
  \city{Antwerp}
  \state{Belgium}
  \postcode{2020}
}

\author{Jeffrey Cornelis}
\email{jeffrey.cornelis@uantwerp.be}
\authornotemark[1]
\affiliation{%
  \institution{University of Antwerp}
  \streetaddress{Middelheimlaan 1}
  \city{Antwerp}
  \state{Belgium}
  \postcode{2020}
}

\author{Pieter Ghysels}
\email{pghysels@lbl.gov}
\authornote{Lawrence Berkeley National Laboratory, Computational Research  Division, 1 Cyclotron Road, Berkeley, CA94720, USA}
\affiliation{%
  \institution{Lawrence Berkeley National Laboratory}
  \streetaddress{1 Cyclotron Road} 
  \city{Berkeley}
  \state{CA, USA}
  \postcode{CA94720}
}

\author{Wim Vanroose}
\email{wim.vanroose@uantwerp.be}
\authornotemark[1]
\affiliation{%
  \institution{University of Antwerp}
  \streetaddress{Middelheimlaan 1}
  \city{Antwerp}
  \state{Belgium}
  \postcode{2020}
}

%
\renewcommand{\shortauthors}{Cools et al.}

%
\begin{abstract}
This paper presents performance results comparing MPI-based implementations of the popular Conjugate Gradient (CG) method and several of its communication hiding (or ``pipelined'') variants. Pipelined CG methods are designed to efficiently solve SPD linear systems on massively parallel distributed memory hardware, and typically display significantly improved strong scaling compared to classic CG. This increase in parallel performance is achieved by overlapping the global reduction phase (\texttt{MPI\_Iallreduce}) required to compute the inner products in each iteration by (chiefly local) computational work such as the matrix-vector product as well as other global communication. 
This work includes a brief introduction to the deep pipelined CG method for readers that may be unfamiliar with the specifics of the method. A brief overview of implementation details provides 
the practical tools required for implementation of the algorithm. Subsequently, easily reproducible strong scaling results on the US Department of Energy (DoE) NERSC machine ``Cori'' (Phase I -- Haswell nodes) on up to 1024 nodes with 16 MPI ranks per node are presented using an implementation of p($l$)-CG that is available in the open source PETSc library. 
Observations on the staggering and overlap of the asynchronous, non-blocking global communication phases with communication and computational kernels are drawn from the experiments. 
\end{abstract}

%
%


%

%
\begin{teaserfigure}
\centering
		\includegraphics[width=0.85\linewidth]{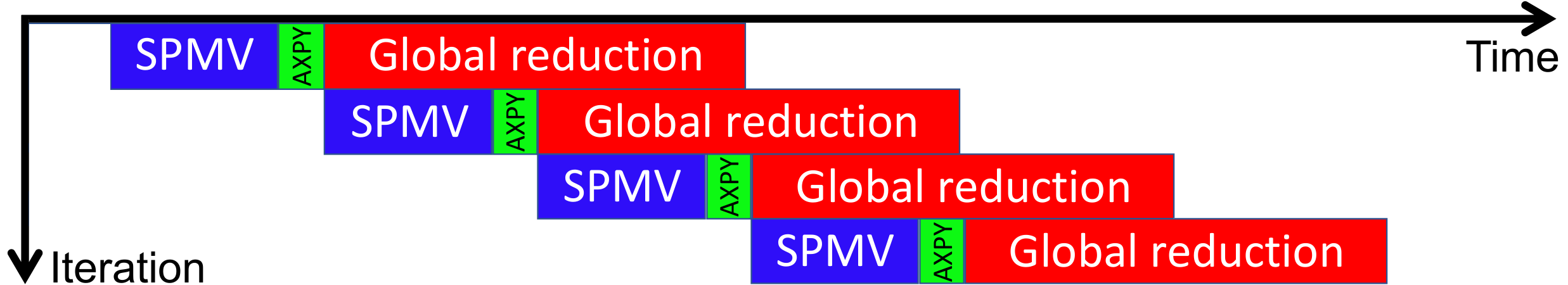}
		\hfill
		\caption{Schematic representation of global reduction pipelining in Krylov subspace methods (e.g.~Conjugate Gradients)
		for pipeline length two ($l = 2$). Global communication is initiated by an MPI$\_$Iallreduce call. The reduction overlaps with the global communication and computational kernels in the next two iterations and is finalized by MPI$\_$Wait. Optimally, a theoretical $\mathcal{O}(l)$ speedup over classic Krylov subspace methods is achieved.}
	\label{fig:teaser}
	\vspace{0.5cm}
\end{teaserfigure}

%
\maketitle

\section{Introduction}

The family of iterative solvers known as Krylov subspace methods (KSMs) \cite{greenbaum1997iterative,liesen2012krylov,saad2003iterative,van2003iterative} are among the most efficient present-day methods for solving large scale sparse systems of linear equations. The Conjugate Gradient method (CG) can be considered as the mother of all Krylov subspace methods. It was 
derived in 1952 \cite{hestenes1952methods} to the aim of solving linear systems $Ax=b$ with a sparse symmetric positive definite (SPD) matrix $A$. The CG method is one of the most widely used methods for solving these systems today, which in turn form the basis of a plethora of scientific and industrial applications. However, driven by the 
essential transition from optimal time-to-solution on a single node towards massively parallel computing over the last decades \cite{shalf2007new}, the bottleneck for fast execution of Krylov subspace methods has shifted. Whereas in the past the application of the sparse matrix-vector product (\textsc{spmv}) was considered the most time-consuming part of the algorithm, the global synchronizations required in dot product and norm computations form the main bottleneck for efficient execution on present-day distributed memory hardware \cite{dongarra2011international,dongarra2015hpcg}.

Driven by the increasing levels of parallelism in present-day HPC machines, research on the elimination of the global communication bottleneck has recently regained significant attention from the international computer science, engineering and numerical mathematics communities.
Evolving from pioneering work on reducing communication in Krylov subspace methods from the late 1980's and 90's \cite{strakovs1987effectivity,meurant1987multitasking,chronopoulos1989s,d1992reducing,demmel1993parallel,erhel1995parallel
}, a number of variants of the classic Krylov subspace algorithms have been introduced over the last years. We specifically point out recent work by Chronopoulos et al.~\cite{chronopoulos2010block}, Hoemmen \cite{hoemmen2010communication}, Carson et al.~\cite{carson2013avoiding,carson2014residual}, McInnes et al.~\cite{mcinnes2014hierarchical}, Grigori et al.~\cite{grigori2016enlarged}, Eller et al.~\cite{eller2016scalable}, Imberti et al.~\cite{imberti2017varying} and Zhuang et al.~\cite{zhuang2017iteration}.

The research presented in this paper is situated in the research branch on so-called ``\emph{communication hiding}'' or ``\emph{pipelined}'' Krylov subspace methods\footnote{\textbf{Note:} In the context of communication reduction in Krylov subspace methods, the terminology \emph{``pipelined''} KSMs that is used throughout the related applied linear algebra and computer science literature refers to \emph{software pipelining}, i.e. algorithmic reformulations to the KSM procedure in order to reduce communication overhead, and should not be confused with hardware-level \emph{instruction pipelining} (ILP).} \cite{ghysels2013hiding,ghysels2014hiding,cools2017communication}. These algorithmic variations to classic Krylov subspace methods are designed to overlap the time-consuming global communications in each iteration of the algorithm with computational tasks such as calculating \textsc{spmv}s or \textsc{axpy}s (vector operations of the form $y \leftarrow \alpha x + y$) as well as with global communication phases initiated in later iterations. Thanks to the reduction/elimination of the synchronization bottleneck, pipelined algorithms have been shown to \emph{increase parallel efficiency} 
by allowing the algorithm to continue scaling on large numbers of processors \cite{sanan2016pipelined,yamazaki2017improving}. However, when redesigning these algorithms one should be wary not to deteriorate the mathematical properties (numerical stability, attainable accuracy, etc.) that are well established for the original Krylov algorithms. Research on analyzing and improving the numerical stability of pipelined Krylov subspace methods has recently been performed in \cite{carson2018numerical,cools2018analyzing,cools2018analyzing2,cools2019numerically}; we point out the references therein for more information on the numerical properties of Krylov subspace methods.  

Strong scaling results obtained from an MPI-based implementation of the $l$-length pipelined Conjugate Gradient method, p($l$)-CG for short, are presented in this work. The p($l$)-CG method was recently 
presented in \cite{cornelis2017communication,cools2019numerically} and allows to overlap each global reduction phase with the communication and computational work of $l$ subsequent iterations. This overlap is achieved by exploiting the asynchronous non-blocking global ``Iallreduce'' MPI operations that have been available from the MPI3 standard onward. The pipeline length $l$ is a parameter of the method that can be chosen depending on the problem and hardware setup (as a function of the communication-to-computation ratio). As is the case for all communication reducing Krylov subspace methods, a number of problem- and hardware-dependent factors influence the communication-to-computation ratio and thus directly affect the performance of the method. The choice of the pipeline length is subject to the speed of the communication network, the time consumed by computing the matrix-vector product and the effort one is willing to invest in a preconditioner. The argument for a longer pipeline use case is stronger for preconditioners that use limited communication (e.g.~block Jacobi, no-overlap DDM, \ldots).
 
This work aims to provide the reader with insights on the implementation and use of the pipelined Conjugate Gradient method as a promising contemporary application of MPI. The focus of the paper is therefore on providing a brief overview of the properties of the p($l$)-CG method which are required for a practical MPI-based implementation. 


We conclude this introduction by presenting a short overview of the contents of this paper. 
Section \ref{sec:pipelcg} familiarizes the reader with the fundamentals of communication hiding Krylov subspace methods. It presents a high-level summary of the mathematical principles behind the $l$-length pipelined CG method and formulates technical comments on the algorithm's pseudo-code, which is included in this section.  
Section \ref{sec:imple} provides the practical framework for an MPI-based implementation of the method described in Section \ref{sec:pipelcg}. It includes discussions on the ability to overlap the main algorithmic kernels in the algorithm using asynchronous non-blocking \texttt{MPI\_Iallreduce} calls, the computational cost of these kernels, and an overview of the algorithm's storage requirements.
Section \ref{sec:strong} presents various strong scaling results that were obtained using an MPI-based implementation of the pipelined CG algorithm in the PETSc library \cite{petsc-web-page}. The experiments were executed on up to 1024 nodes of the US Department of Energy NERSC facility ``Cori''. The experimental results are supplemented by an elaborate discussion that aims to provide the reader with detailed insights on the performance gains reported in the strong scaling experiments.
The manuscript is concluded by a short outlook in Section \ref{sec:conclusions}.

\section{The pipelined CG algorithm} \label{sec:pipelcg}

The deep pipelined Conjugate Gradient method, denoted p($l$)-CG for short, was first presented in \cite{cornelis2017communication}, where it was derived in analogy to the p($l$)-GMRES method \cite{ghysels2013hiding}. The parameter $l$ 
represents the pipeline length which indicates the number of iterations that are overlapped by each global reduction phase. We summarize the current state-of-the-art deep pipelined p($l$)-CG method below.

\subsection{Mathematical framework for p($l$)-CG} \label{sec:mathematical}

Let $V_{i-l+1}=[v_{0},v_{1},\ldots,v_{i-l}]$ be the orthonormal basis for the Krylov subspace $\mathcal{K}_{i-l+1}(A,v_{0})$ in iteration $i$ of the p($l$)-CG algorithm, consisting of $i-l+1$ vectors. Here $A$ is assumed to be a symmetric positive definite matrix. The Krylov subspace basis vectors satisfy the Lanczos relation
\begin{equation} \label{eq:Arnoldi_v}
AV_{j} = V_{j+1} T_{j+1,j}, \qquad 1 \leq j \leq i-l.
\end{equation}
with \vspace{-0.3cm}
\begin{equation} \label{eq:Arnoldi_v2}
T_{j+1,j} =  
	\begin{pmatrix} 
  \gamma_{0} & \delta_{0} & & \\ 
 \delta_{0} & \gamma_{1} & \ddots & \\  
  & \ddots & \ddots & \delta_{j-2} \\ 
  & & \delta_{j-2} & \gamma_{j-1}  \\
  & & & \delta_{j-1}
	\end{pmatrix}.
\end{equation}
Let $\delta_{-1} = 0$, then the Lanczos relation \eqref{eq:Arnoldi_v} translates in vector notation to 
\begin{equation}\label{eq:v_ex}
  v_{j+1}= ( Av_{j} - \gamma_j v_j - \delta_{j-1} v_{j-1} ) / \delta_j, \quad 0 \leq j < i - l.
\end{equation}
The auxiliary basis $Z_{i+1}=[z_{0},z_{1},\ldots,z_{i}]$ runs $l$ vectors ahead of the basis $V_{i-l+1}$ and is defined as 
\begin{equation} \label{eq:z_ex}
  z_{j}= \left\{ \begin{matrix} v_{0}, & j=0, \\ P_{j}(A)v_{0}, & 0<j\leq l, \\ P_{l}(A)v_{j-l}, & l<j\leq i, \end{matrix} \right. 
\end{equation}
where the matrix polynomial $P_{j}(A)$ is given by
\begin{equation} \label{eq:poly}
	P_{j}(t) = \prod_{k=0}^{j-1} (t-\sigma_{k}), \qquad j\leq l,
\end{equation}
with optional 
stabilizing shifts $\sigma_{k}\in \mathbb{R}$, see \cite{ghysels2013hiding,cornelis2017communication,hoemmen2010communication}. 
The choice of the polynomial $P_l(A)$ is discussed in Section \ref{sec:technical}.
Contrary to the basis $V_{i-l+1}$, the auxiliary basis $Z_{i+1}$ is not orthonormal.
It is constructed using the recursive definitions
\begin{equation} \label{eq:z_rec}
  z_{j+1} = \left\{
	\begin{matrix}
			(A-\sigma_{j}I) \, z_{j}, & 0 \leq j < l, \\
			(Az_{j} - \gamma_{j-l} z_j - \delta_{j-l-1} z_{j-1} ) / \delta_{j-l}, & l \leq j < i, 
	\end{matrix} \right.
\end{equation}
which are obtained by left-multiplying the Lanczos relation \eqref{eq:v_ex} on both sides by $P_{l}(A)$.
Expression \eqref{eq:z_rec} translates into a Lanczos type matrix relation
\begin{equation}
A Z_j = Z_{j+1} B_{j+1,j}, \qquad 1 \leq j \leq i,
\end{equation}
where the matrix $B_{j+1,j}$ contains the matrix $T_{j-l+1,j-l}$, which is shifted $l$ places along the main diagonal, i.e.
\begin{equation}
B_{j+1,j} = 
\left(\begin{array}{ccc|ccc} 
\sigma_{0} & & & & & \\
1 & \ddots & & & & \\
& \ddots & \sigma_{l-1} & & & \\ \hline
& & 1 & & & \\
& & & & T_{j-l+1,j-l} & \\
& & & & & 
\end{array}\right).
\end{equation}
The bases $V_{j}$ and $Z_j$ are connected through the basis transformation $Z_{j}=V_{j}G_{j}$ for $1 \leq j \leq i-l+1$, where $G_j$ is a banded upper triangular matrix with a band width of $2l+1$ non-zero diagonals \cite{cornelis2017communication}. For a symmetric matrix $A$ the matrix $G_{i+1}$ is symmetric around its $l$-th upper diagonal, since
\begin{align} \label{eq:symmetry_G}
	g_{j,i} &= (z_{i},v_{j}) = (P_{l}(A)v_{i-l},v_{j}) = (v_{i-l},P_{l}(A)v_{j}) \notag \\ 
					&= (v_{i-l},z_{j+l}) = g_{i-l,j+l}.
\end{align}
The following recurrence relation for $v_{j+1}$ is derived from the basis transformation (with $0 \leq j < i-l$):
\begin{equation} \label{eq:v_rec}
  v_{j+1} = \left( z_{j+1} - \sum_{k=j-2l+1}^{j} g_{k,j+1} v_{k} \right) / g_{j+1,j+1}.
\end{equation}
A total of $l$ iterations after the dot-products 
\begin{equation}
g_{j,i+1} = \left\{ \begin{matrix}
	(z_{i+1},v_{j}); & j=i-2l+1,\ldots,i-l+1, \\ 
	(z_{i+1},z_{j});  &j=i-l+2,\ldots,i+1, \end{matrix}
	\right.
\end{equation}
have been initiated, the elements $g_{j,i-l+1}$ with $i-2l+2 \leq j \leq i-l +1$, which were computed as $(z_{i-l+1},z_{j})$, are corrected as follows:
\begin{equation}
g_{j,i-l+1} = \frac{g_{j,i-l+1}-\sum_{k=i-3l+1}^{j-1}g_{k,j}g_{k,i-l+1}}{g_{j,j}}, 
\end{equation}
for $j=i-2l+2,\ldots, i-l$, and
\begin{equation}
g_{i-l+1,i-l+1} =\sqrt{g_{i-l+1,i-l+1}-\sum_{k=i-3l+1}^{i-l}g_{k,i-l+1}^2}.
\end{equation}
The tridiagonal matrix $T_{i-l+2,i-l+1}$, see \eqref{eq:Arnoldi_v2}, is updated recursively by adding one column in each iteration $i$. 
The diagonal element $\gamma_{i-l}$ is characterized by the expressions: $\gamma_{i-l} = $
\begin{equation} \label{eq:haa1}
\left\{
  \begin{aligned} 
    & \frac{g_{i-l,i-l+1}+\sigma_{i-l}g_{i-l,i-l} -g_{i-l-1,i-l}\delta_{i-l-1}}{g_{i-l,i-l}}, \quad l \leq i < 2l,\\
    & \frac{g_{i-l,i-l}\gamma_{i-2l}+g_{i-l,i-l+1}\delta_{i-2l}-g_{i-l-1,i-l}\delta_{i-l-1}}{g_{i-l,i-l}}, ~i \geq 2l.
  \end{aligned} 
	\right.
\end{equation}
The term $-g_{i-l-1,i-l}\delta_{i-l-1}$
is considered zero when $i = l$.
The update for the off-diagonal element $\delta_{i-l}$ is given by
\begin{equation} \label{eq:haa3}
\delta_{i-l} = \left\{ 
  \begin{matrix} 
    g_{i-l+1,i-l+1}/g_{i-l,i-l}, & l \leq i < 2l, \\ 
    (g_{i-l+1,i-l+1}\delta_{i-2l})/g_{i-l,i-l}, & i \geq 2l.
  \end{matrix} 
	\right.
\end{equation}
The element $\delta_{i-l-1}$ has already been computed in the previous iteration and can thus simply be copied due to the symmetry of $T_{i-l+2,i-l+1}$.

\begin{algorithm*}[t]
{\small
\caption{Preconditioned $l$-length pipelined p($l$)-CG \hfill 
\textbf{Input:} $A$, $M^{-1}$, $b$, $x_0$, $l$, $m$, $\tau$, $\left\{\sigma_0,\ldots,\sigma_{l-1}\right\}$}\label{algo:PIPELCG}
\begin{algorithmic}[1]
\State $u_{0}:=b-Ax_{0};$ $r_{0}:=M^{-1}u_{0};$ $z^{(l)}_{0}:= z^{(l-1)}_{0}:= \,\ldots\, := z^{(1)}_{0}:=z^{(0)}_{0} := r_{0}/\sqrt{(u_0,r_0)};  ~ g_{0,0}:=1;$
\For {$i=0,\ldots, m+l$}
\State $u_{i+1}:=\left\{ \begin{matrix}Az^{(l)}_{i}-\sigma_{i}u_{i}, & i<l \\ Az^{(l)}_{i}, & i \geq l \end{matrix}\right.$ \hfill \# Compute matrix-vector product
\State $z^{(l)}_{i+1}:=M^{-1}u_{i+1}$; \hfill \# Apply preconditioner
\If {$i<l-1$}
\State $z^{(k)}_{i+1} := z^{(l)}_{i+1}; \qquad k = i+1 ,\ldots, l-1$ \hfill \text{\# Copy auxiliary basis vectors} 
\EndIf
\State \textbf{end if}
\If {$i\geq l$} \hfill \# Finalize dot-products ($g_{j,i-l+1}$)
\State $g_{j,i-l+1} := (g_{j,i-l+1}-\sum_{k=i-3l+1}^{j-1}g_{k,j}g_{k,i-l+1})/g_{j,j}; \qquad j=i-2l+2,\ldots,i-l$ \hfill \# Update transformation matrix
\State $g_{i-l+1,i-l+1}:= \sqrt{g_{i-l+1,i-l+1}-\sum_{k=i-3l+1}^{i-l}g_{k,i-l+1}^2};$
\State \# Check for breakdown and restart if required \hfill \# Square root breakdown check
\If {$i<2l$}
\State $\gamma_{i-l}:=(g_{i-l,i-l+1}+\sigma_{i-l}g_{i-l,i-l} - g_{i-l-1,i-l}\delta_{i-l-1})/g_{i-l,i-l};$ \hfill \# Add column to Hessenberg matrix
\State $\delta_{i-l}:=g_{i-l+1,i-l+1}/g_{i-l,i-l};$
\Else
\State $\gamma_{i-l}:=(g_{i-l,i-l}\gamma_{i-2l}+g_{i-l,i-l+1}\delta_{i-2l} -g_{i-l-1,i-l}\delta_{i-l-1})/g_{i-l,i-l};$
\State $\delta_{i-l}:=(g_{i-l+1,i-l+1}\delta_{i-2l})/g_{i-l,i-l};$
\EndIf
\State \textbf{end if}
\State $z^{(k)}_{i-l+k+1} := (z^{(k+1)}_{i-l+k+1} + (\sigma_k - \gamma_{i-l}) z^{(k)}_{i-l+k} - \delta_{i-l-1} z^{(k)}_{i-l+k-1})/\delta_{i-l}; \qquad k = 0, \ldots, l-1$ \hfill \# Compute basis vectors
\State $z^{(l)}_{i+1} := (z^{(l)}_{i+1} - \gamma_{i-l} z^{(l)}_{i}- \delta_{i-l-1} z^{(l)}_{i-1})/\delta_{i-l};$ \hfill\# Compute basis vector
\State $u_{i+1} := (u_{i+1} - \gamma_{i-l} u_{i}- \delta_{i-l-1} u_{i-1})/\delta_{i-l};$ \hfill\# Compute basis vector
\EndIf
\State \textbf{end if}
\State $g_{j,i+1}:=\left\{ \begin{matrix}(u_{i+1},z^{(0)}_{j}); & j=\max(0,i-2l+1),\ldots,i-l+1 \\ (u_{i+1},z^{(l)}_{j});  &j=i-l+2,\ldots,i+1 \end{matrix}\right.$ \hfill\# Initiate dot-products ($g_{j,i+1}$)
\If {$i=l$}
\State $\eta_{0}:=\gamma_{0};$ \quad $\zeta_{0}:=\sqrt{(u_0,r_0)};$ \quad $p_{0}:=z^{(0)}_0/\eta_0;$
\Else \, \textbf{if} {$i\geq l+1$} \textbf{then}
\State $\lambda_{i-l}:=\delta_{i-l-1}/\eta_{i-l-1};$ \hfill \# Factorize Hessenberg matrix
\State $\eta_{i-l}:=\gamma_{i-l}-\lambda_{i-l}\delta_{i-l-1};$ 
\State $\zeta_{i-l}=-\lambda_{i-l}\zeta_{i-l-1};$ \hfill \# Compute recursive residual norm
\State $p_{i-l}=(z^{(0)}_{i-l}-\delta_{i-l-1}p_{i-l-1})/\eta_{i-l};$ \hfill \# Update search direction
\State $x_{i-l}=x_{i-l-1}+\zeta_{i-l-1}p_{i-l-1};$ \hfill \# Update approximate solution
\If {$|\zeta_{i-l}|/\sqrt{(u_0,r_0)} < \tau$} RETURN; \textbf{end if}  \hfill \# Check convergence criterion
\EndIf
\EndIf
\State \textbf{end if}
\EndFor 
\end{algorithmic}
}
\end{algorithm*}

Once the basis $V_{i-l+1}$ has been constructed, the solution $x_{i-l}$ can be updated based on a search direction $p_{i-l}$, following the classic derivation of D-Lanczos in \cite{saad2003iterative}, Sec.~6.7.1. The Ritz-Galerkin condition (that holds for $1 \leq j \leq i-l+1$)
\begin{align} \label{eq:lanczos}
  0 &= V_j^T r_j  = V_j^T (r_0 - A V_j y_j) \notag \\
		&= V^T_j \left(V_j {\|r_0\|}_2 e_1\right) - T_j y_j = {\|r_0\|}_2 e_1 - T_j y_j,
\end{align}
implies $y_j = T_{j}^{-1} {\|r_0\|}_2 e_1$.
The LU-factorization of the tridiagonal matrix $T_{i-l+1} = L_{i-l+1} U_{i-l+1}$ is given by
\begin{equation} \label{eq:LU}  
\begin{pmatrix} 
1 & & & \\ 
\lambda_{1} & 1 & &  \\  
& \ddots & \ddots &  \\ 
& & \lambda_{i-l} & 1  
\end{pmatrix}
\begin{pmatrix} 
\eta_{0} & \delta_{0} & & \\ 
& \eta_{1} & \ddots & \\  
& & \ddots & \delta_{i-l-1} \\ 
& & & \eta_{i-l} 
\end{pmatrix}.
\end{equation}
Note that $\gamma_{0}=\eta_{0}$. It follows from \eqref{eq:LU} that the elements of the lower/upper triangular matrices $L_{i-l+1}$ and $U_{i-l+1}$ are given by (with $1 \leq j \leq i-l$)
\begin{equation}
\lambda_{j}=\delta_{j-1}/\eta_{j-1} \quad \text{and} \quad \eta_{j}= \gamma_{j}-\lambda_{j}\delta_{j-1}. 
\end{equation}
Expression \eqref{eq:lanczos} indicates that the approximate solution $x_{i-l+1}$ equals 
\begin{equation} \label{eq:x_a}
x_{i-l+1} = x_{0}+P_{i-l+1} q_{i-l+1},
\end{equation}
where $P_{i-l+1} = V_{i-l+1} U_{i-l+1}^{-1}$ and $q_{i-l+1} = L_{i-l+1}^{-1}\|r_{0}\|_{2}e_{1}$. It holds that $p_0 = v_0/\eta_0$. The columns $p_j$ (for $1 \leq j \leq i-l$) of $P_{i-l+1}$ can be computed recursively. From $P_{i-l+1} U_{i-l+1} = V_{i-l+1}$ it follows 
\begin{equation}\label{eq:pdir}
p_j = \eta_j^{-1}(v_j-\delta_{j-1} p_{j-1}), \qquad 1 \leq j \leq i-l.
\end{equation}
Denoting the 
vector $q_{i-l+1}$ by $\left[\zeta_0,\ldots,\zeta_{i-l}\right]^T$, it follows from 
\begin{equation}
L_{i-l+1} q_{i-l+1} = \|r_{0}\|_{2} e_{1} 
\end{equation}
that $\zeta_0 = {\|r_0\|}_2$ and $\zeta_j = -\lambda_{j}\zeta_{j-1}$ for $1\leq j \leq i-l$. 
The approximate solution $x_{i-l}$ is then updated using the recurrence relation
\begin{equation} \label{eq:x_a2}
x_{i-l} =x_{i-l-1} + \zeta_{i-l-1}p_{i-l-1}.
\end{equation}
The above expressions are combined in Alg.\,\ref{algo:PIPELCG}. 
Once the initial pipeline for $z_0,\ldots,z_l$ has been filled, the relations \eqref{eq:z_rec}-\eqref{eq:v_rec} can be used 
to recursively compute the basis vectors $v_{i-l+1}$ and $z_{i+1}$ in iterations $i \geq l$ (see lines 19-21).
The scalar results of the global reduction phase 
(line 23) are required $l$ iterations later (lines 9-10). In every iteration global communication is thus overlapped with the computational work and the communications of $l$ subsequent iterations. This \emph{``communication hiding''} forms the core property of the p($l$)-CG algorithm.
Note that Alg.\,\ref{algo:PIPELCG} presents a complete version of the algorithm including preconditioning, stable recurrences, stopping criteria, etc. These topics are covered in more detail in Section \ref{sec:technical} below.

\subsection{Technical comments on the algorithm}\label{sec:technical}
To complement the mathematical framework a number of insightful technicalities of the algorithm are discussed.

\paragraph{\textbf{Preconditioned p($l$)-CG}}
Alg.\,\ref{algo:PIPELCG} is written in a general form that also includes preconditioning. This requires to compute the unpreconditioned auxiliary variables $u_{i}=M z_{i}$ in addition to the quantities described in Section \ref{sec:mathematical}. These variables also satisfy a Lanczos type relation: 
\begin{equation} \label{eq:zhatid3}
u_{i+1} 
= \left\{ \begin{matrix}Az_{i} -\sigma_{i}u_{i}  & i<l, \\ (Az_{i} - \gamma_{i-l} u_{i}- \delta_{i-l-1} u_{i-1})/\delta_{i-l} & i \geq l,  \end{matrix}\right.
\end{equation} 
The corresponding dot-products on line 23 are defined as $M$-inner products in the context of the preconditioned algorithm, see \cite{cools2019numerically} for details.

\paragraph{\textbf{Residual norm in p($l$)-CG}}
The residual $r_j = b-Ax_j$ is not computed in Alg.\,\ref{algo:PIPELCG}, but its norm is characterized by the quantity $|\zeta_j| = \|r_j\|_2$ (or its natural norm equivalent $|\zeta_j| = \sqrt{(u_0,r_0)}$ in the preconditioned case) for $0 \leq j \leq i-l$. 
This quantity is useful to formulate a stopping criterion for the p($l$)-CG iteration without introducing additional communication or computations, see Alg.\,\ref{algo:PIPELCG}, line 32.

\paragraph{\textbf{Number of dot products in p($l$)-CG}}
Although Alg.\,\ref{algo:PIPELCG}, line 23, indicates that in each iteration $i \geq 2l+1$ a total of $(2l+1)$ dot products need to be computed, the number of dot product computations can be limited to $(l+1)$ by exploiting the symmetry of the matrix $G_{i+1}$, see expression \eqref{eq:symmetry_G}. Since $g_{j,i+1} = g_{i-l+1,j+l}$ for $j \leq i+1$, only the dot products $(u_{i+1},z_{j})$ for $j=i-l+2,\ldots,i+1$
and the $l$-th upper diagonal element $(u_{i+1},v_{i-l+1})$ need to be computed. 



\paragraph{\textbf{Choice of auxiliary basis shifts}}
The auxiliary basis vectors $z_j$ are defined as $P_l(A) v_{j-l}$, but the basis $Z_{i}$ is in general not orthogonal. For longer pipelined lengths $l$, $Z_i^T Z_i$ may be very ill-conditioned and may become close to singular as $i$ augments. 
The effect is the most pronounced when $\sigma_0 = \ldots = \sigma_{l-1} = 0$, in which case $P_l(A) = A^l$. Shifts $\sigma_j$ can be set to improve the conditioning of $Z_i^T Z_i$. It holds that
\begin{equation}
	Z_{l+1:i}^T Z_{l+1:i} = G_{l+1:i}^T G_{l+1:i} = V_{i-l}^T \, P_l(A)^2 \, V_{i-l},
\end{equation}
where $Z_{l+1:i}$ is a part of the basis $Z_i$ obtained by dropping the first $l$ columns.
Hence, the polynomial $P_l(A)^2$ has a major impact on the conditioning of the matrix $G_i^{-1}$, which in turn plays a crucial role in the propagation of local rounding errors in the p($l$)-CG algorithm \cite{cornelis2017communication}. This observation motivates minimization of $\|P_l(A)\|_2$. Optimal shift values for minimizing the Euclidean $2$-norm of $P_l(A)$ are the Chebyshev shifts \cite{hoemmen2010communication,ghysels2013hiding,cornelis2017communication} (for $i=0,\ldots,l-1$):
\begin{equation} \label{eq:chebyshev}
  \sigma_{i} = \frac{\lambda_{\max}+\lambda_{\min}}{2}+\frac{\lambda_{\max}-\lambda_{\min}}{2} \cos\left(\frac{(2i+1)\pi}{2l}\right),
\end{equation}
which are used throughout this work. This requires a notion of the largest (
smallest) eigenvalue $\lambda_{\max}$ (resp.~$\lambda_{\min}$), which can be estimated a priori, e.g.~by 
a few 
power method iterations.

\paragraph{\textbf{Square root breakdowns in p($l$)-CG}}
When for certain $i$ the matrix $Z_i^T Z_i$ becomes (numerically) singular, the Cholesky factorization procedure in p($l$)-CG will fail. This may manifest in the form of a square root breakdown on line 10 in Alg.\,\ref{algo:PIPELCG} when the root argument 
becomes negative. Numerical round-off errors may increase the occurrence of these breakdowns in practice, cf.~\cite{cornelis2017communication,cools2019numerically}. When a breakdown occurs the p($l$)-CG iteration is restarted explicitly. 
This restarting may delay convergence compared to standard CG, where no square root breakdowns occur.

\paragraph{\textbf{Numerically stable recurrences}} 

Although expression \eqref{eq:v_rec} provides a recursive definition for the basis $V_{i+1}$, it was shown in \cite{cornelis2017communication} that this relation may lead to numerical unstable convergence behavior. We therefore introduce a total of $l+1$ bases, denoted by $Z^{(k)}_{i+1}$, where the upper index `$(k)$' ($0 \leq k \leq l$) labels the different bases and the lower index `$i+1$' indicates the iteration. The auxiliary bases are defined as:
\begin{equation} \label{eq:def_zjk}
z^{(k)}_{j} = 
\left\{ 
	\begin{matrix} 
		v_{0}, & j=0, \\ 
		P_{j}(A)v_{0}, & 0<j\leq k, \\ 
		P_{k}(A)v_{j-k}, & k<j\leq i, 
	\end{matrix} 
\right. \quad \text{for}~~ 0 \leq k \leq l,
\end{equation}
where the polynomial is defined by \eqref{eq:poly}. By definition \eqref{eq:def_zjk} the basis $Z^{(0)}_{i+1}$ denotes the original Krylov subspace basis
\begin{equation}
Z^{(0)}_{i+1} = V_{i+1},
\end{equation} 
whereas the $l$-th basis $Z^{(l)}_{j}$ is precisely the auxiliary basis $Z_j$ that was defined earlier, see \eqref{eq:z_ex}, i.e.
\begin{equation}
Z^{(l)}_{i+1} = Z_{i+1},
\end{equation} 
Intuitively, the $k$-th basis $Z^{(k)}_{i+1}$ should be interpreted as running $k$ \textsc{spmv}s ahead of the original Krylov basis $V_{i+1}$.

Note that the first $k+1$ vectors in the basis $Z^{(k)}_{j}$ $(j \geq k)$ and all bases $Z^{(m)}_{j}$ with $m \geq k$ are identical, which is represented by line 6 in Alg.\,\ref{algo:PIPELCG}.
For each pair of subsequent bases $Z^{(k)}_j$ and $Z^{(k+1)}_j$ (for $j > k$) it holds that
\begin{equation} \label{eq:all_z_relation}
  Az^{(k)}_j = z^{(k+1)}_{j+1} + \sigma_{k} z^{(k)}_{j}, \quad j \geq k, \quad 0 \leq k \leq l-1.
\end{equation}
By multiplying the original Lanczos relation \eqref{eq:v_ex} for $v_j$ on both sides by the respective polynomial $P_k(A)$ with $1 \leq k \leq l$ and by exploiting the associativity of $A$ and $P_k(A)$, it is straightforward to see that each auxiliary basis $Z^{(k)}_j$ satisfies a Lanczos type recurrence relation:
\begin{equation} \label{eq:all_z_rec}
z^{(k)}_{j+1} = (A z^{(k)}_{j} - \gamma_{j-k} z^{(k)}_{j} - \delta_{j-k-1} z^{(k)}_{j-1}) / \delta_{j-k}, 
\end{equation}
for $j \geq k$ and $0 \leq k \leq l$.
When $k = 0$ expression \eqref{eq:all_z_rec} yields the Lanczos relation \eqref{eq:v_ex} for $v_{j+1}$, whereas setting $k = l$ results in the recurrence relation \eqref{eq:z_rec} for $z_{j+1}$. 
To avoid the computation of additional \textsc{spmv}s, the relations \eqref{eq:all_z_rec} can be rewritten using expression \eqref{eq:all_z_relation} as:
\begin{equation} \label{eq:all_z_rec2}
z^{(k)}_{j+1} = (z^{(k+1)}_{j+1} + (\sigma_{k} - \gamma_{j-k}) z^{(k)}_{j} - \delta_{j-k-1} z^{(k)}_{j-1}) / \delta_{j-k},
\end{equation}
with $j \geq k$ and $0 \leq k < l$. The recurrence relations \eqref{eq:all_z_rec2} allow to compute the vector updates for the bases $Z^{(0)}_j, \ldots, Z^{(l-1)}_j$ without the need to compute \textsc{spmv}s. The recursive update \eqref{eq:all_z_rec} is used \emph{only} for $k = l$, i.e.~to compute the vectors in the auxiliary basis $Z^{(l)}_j = Z_j$ we use the recurrence relation
\begin{equation} \label{eq:specific_z_rec}
z^{(l)}_{j+1} = (A z^{(l)}_{j} - \gamma_{j-l} z^{(l)}_{j} - \delta_{j-l-1} z^{(l)}_{j-1}) / \delta_{j-l}, 
\end{equation}
for $j \geq l$, which pours down to the recurrence relation \eqref{eq:z_rec}. 

The recursive relations derived above are summarized on lines 19-21 in Alg.\,\ref{algo:PIPELCG}.
In the $i$-th iteration of Alg.\,\ref{algo:PIPELCG} each basis $Z^{(0)}_j, Z^{(1)}_j, \ldots, Z^{(l)}_j$ is updated by adding one vector. The algorithm thus computes a total of $l+1$ new basis vectors, i.e.: $\{v_{i-l+1} = z^{(0)}_{i-l+1}, z^{(1)}_{i-l+2}, \ldots, z^{(l-1)}_{i}, z^{(l)}_{i+1} = z_{i+1}\}$, per iteration. For each basis, the corresponding vector update is
\begin{align*}
\left\{ 
	\begin{array}{lll}
	z^{(0)}_{i-l+1} &\hspace{-0.2cm}= (z^{(1)}_{i-l+1} + (\sigma_{0} - \gamma_{i-l}) z^{(0)}_{i-l} - \delta_{i-l-1} z^{(0)}_{i-l-1}) / \delta_{i-l}, \\ 
	z^{(1)}_{i-l+2} &\hspace{-0.2cm}= (z^{(2)}_{i-l+2} + (\sigma_{1} - \gamma_{i-l}) z^{(1)}_{i-l+1} - \delta_{i-l-1} z^{(1)}_{i-l}) / \delta_{i-l}, \\ 
	  \quad \vdots	&\hspace{-0.2cm} \qquad \qquad\qquad \qquad \qquad	\vdots  \\
	z^{(l-1)}_{i}   &\hspace{-0.2cm}= (z^{(l)}_{i} + (\sigma_{l-1} - \gamma_{i-l}) z^{(l-1)}_{i-1} - \delta_{i-l-1} z^{(l-1)}_{i-2}) / \delta_{i-l}, \\ 
	z^{(l)}_{i+1}   &\hspace{-0.2cm}= (A z^{(l)}_{i} - \gamma_{i-l} z^{(l)}_{i} - \delta_{i-l-1} z^{(l)}_{i-1}) / \delta_{i-l}.
	\end{array}
\right.
\end{align*}
All vector updates make use of the \emph{same} scalar coefficients $\gamma_{i-l}$, $\delta_{i-l}$ and $\delta_{i-l-1}$ that are computed in iteration $i$ of Alg.\,\ref{algo:PIPELCG} (lines 12-18). Only one \textsc{spmv}, namely $A z^{(l)}_{i}$, is required per iteration to compute all basis vector updates.

\section{MPI-based implementation} \label{sec:imple}

Section \ref{sec:pipelcg} gave an overview of the mathematical properties of the p($l$)-CG method. 
In this section we comment on technical aspects concerning the implementation of Alg.\,\ref{algo:PIPELCG}.

\subsection{Hiding global communication 
} \label{sec:hiding}

A schematic kernel-based representation of the p($l$)-CG algorithm is introduced in this section. 
The following computational kernels are defined in iteration $i$ in Alg.\,\ref{algo:PIPELCG}:

\vspace{0.2cm}
\begin{center}
  \begin{tabular}{| c | c | l | c | }
    \hline
    tag & type & kernel description & Alg.\,\ref{algo:PIPELCG} line  \\ \hline \hline
    (K1) & \textsc{spmv}   & apply $A$ and $M^{-1}$ 								& 3-4 \\ \hline
    (K2) & \textsc{scalar} & update matrix $G_{i-l+2}$       				& 9-10 \\ \hline
		(K3) & \textsc{scalar} & update matrix $T_{i-l+2,i-l+1}$  			& 12-18 \\ \hline
		(K4) & \textsc{axpy}   & recursive basis updates       					& 19-21 \\ \hline
		(K5) & \textsc{dotpr}  & compute dot products 									& 23 \\ \hline
		(K6) & \textsc{axpy}   & update $x_{i-l}$ and $|\zeta_{i-l}|$ 	& 24-33 \\ \hline
  \end{tabular}
\end{center}\ 
\vspace{0.2cm}

\noindent The \textsc{spmv} kernel (K1) is considered to be the most computationally intensive part of the algorithm, which is overlapped with the global reduction phase in (K5) to hide communication latency and idle core time. Kernels (K2), (K3), (K4) and (K6) represent local operations which are assumed to have a very low arithmetic complexity when executed on large scale multi-node hardware. These operations are also overlapped with the global reduction phase in p($l$)-CG for pipeline lengths $l>1$. 
In (K5) all local contributions to the dot products are computed by each worker and subsequently a global reduction phase is performed. The preconditioned p($l$)-CG algorithm can be summarized schematically using these kernel definitions as displayed in Alg.\,\ref{algo:schema}. 

\begin{algorithm}[t]
{\small
\caption{Schematic representation of p($l$)-CG \hfill}\label{algo:schema}
\begin{algorithmic}[1]
\State \textsc{initialization} ;
\For {$i=0,\ldots, m+l$}
\State (K1) \textsc{spmv} ;
\If {$i\geq l$}
\State \texttt{\color{red}MPI\_Wait(req(i-l),\ldots)} ;
\State (K2) \textsc{scalar} ;
\State (K3) \textsc{scalar} ;
\State (K4) \textsc{axpy} ;
\EndIf
\State \textbf{end if}
\State (K5) \textsc{dotpr} ;
\State \texttt{\color{red} MPI\_Iallreduce(\ldots,G(i-2l+1:i+1,i+1),\ldots,req(i))} ;
\State (K6) \textsc{axpy} ;
\EndFor 
\State \textbf{end for}
\State \textsc{finalization (drain the pipeline)} ;
\end{algorithmic}
}
\end{algorithm}

Our implementation of Alg.\,\ref{algo:PIPELCG} uses the MPI-3 standard communication library, which allows for asynchronous progress in the reduction by setting the environment variables
\begin{verbatim}
  MPICH_ASYNC_PROGRESS=1;
  MPICH_MAX_THREAD_SAFETY=multiple;
\end{verbatim} 
Note that we also allow for asynchronous communication when running the classic CG method in our experiments.
Global communication is initiated by an \texttt{MPI$\_$Iallreduce} call which starts a non-blocking reduction:
\begin{verbatim}
  MPI_Iallreduce(...,G(i-2l+1:i+1,i+1),...,req(i));
\end{verbatim} 
The argument \texttt{G(i-2l+1:i+1,i+1)} represents the $2l+1$ elements of the matrix $G_{i+2}$ that are computed in (K5) in iteration $i$. The result of the corresponding global reduction phase is signaled to be due to arrive by the call to
\begin{verbatim}
  MPI_Wait(req(i),...);
\end{verbatim} 
The \texttt{MPI\_Request} array element \texttt{req(i)} that is passed to \texttt{MPI\_Wait} keeps track of the iteration index in which the global reduction phase was initiated.
Since the p($l$)-CG method overlaps $l$ \textsc{spmv}'s with a single global reduction phase, the call to 
\begin{verbatim}
  MPI_Wait(req(i),...); 
\end{verbatim}
occurs in iteration $i+l$, i.e.\,$l$ iterations after the call 
\begin{verbatim}
  MPI_Iallreduce(..., req(i)).
\end{verbatim}
The schematic representation, Alg.\,\ref{algo:schema}, shows that the global reduction phase that is initiated by \texttt{MPI\_Iallreduce} with request \texttt{req(i)} in iteration $i$ overlaps with a total of $l$ \textsc{spmv}'s, namely the kernels (K1) in iterations $i+1$ up to $i+l$. The corresponding call to \texttt{MPI\_Wait} with request \texttt{req((i+l)-l)} = \texttt{req(i)} takes place in iteration $i+l$ before the computations of (K2) in which the dot product results are required, but after the \textsc{spmv} kernel (K1) has been executed. The global reduction also overlaps with the less computationally intensive operations (K2), (K3), (K4) and (K6). Hence, 
the global communication latency of the dot products in (K5) is `hidden' behind the computational work of $l$ p($l$)-CG iterations.

Note that apart from overlapping global communication with computational kernels as described above, the p($l$)-CG method also overlaps the global communication phase with other global communication phases (namely from the next $l-1$ iterations) when the pipeline length $l$ is larger than $1$. Additional insights on this topic can be found in Section \ref{sec:discussion}.

\subsection{Computational and storage costs } \label{sec:computational}
 
Table \ref{tab:pipelcg} gives an overview of implementation details of the p($l$)-CG method, Alg.\,\ref{algo:PIPELCG}, including storage requirements and number of \emph{flops} (floating point operations) per iteration. Preconditioning is excluded here for simplicity. We compare to the same properties for Ghysels' p-CG method \cite{ghysels2014hiding}. 
The latter algorithm is conceptually equivalent to p($1$)-CG but was derived in an essentially different way, see \cite{ghysels2014hiding,cools2017communication,cools2018analyzing}. 

\begin{table}[t]
\centering
\small
\caption{\footnotesize Specifications of different CG variants (without preconditioner); 
`CG' denotes classic CG; `p-CG' is Ghysels' pipelined CG \cite{ghysels2014hiding}.
Columns \emph{GLRED} and \emph{SPMV} list the number of 
synchronization phases and \textsc{SPMV}s per iteration. 
The column \emph{Flops} indicates the number of flops ($\times N$) required to compute 
\textsc{AXPY}s and dot products (with $l \geq 1$). The \emph{Time} column shows the 
time spent in global all-reduce communications and \textsc{SPMV}s. 
\emph{Memory} counts the number of vectors 
in memory (excl.~$x_{i-l}$ and $b$) at any time
during 
execution. 
}
\begin{tabular}{| l | c | c | c | c | c |}
\hline 
																			& \rotatebox[origin=c]{90}{~\textsc{glred}~}
																			& \rotatebox[origin=c]{90}{\textsc{spmv}}
																			& \rotatebox[origin=c]{00}{Flops} 
																			& \rotatebox[origin=c]{00}{Time} 
																			& \rotatebox[origin=c]{00}{Memory}\\
\hline 
CG	 																	& 2 & 1 		& 10			& 2 \textsc{glred} + 1 \textsc{spmv} & 3 \\
p-CG																	& 1 & 1     & 16 			& $\max$(\textsc{glred}, \textsc{spmv}) & 6 \\
p($l$)-CG &	1 & 1     & \hspace{-0.2cm} $6l+10$ \hspace{-0.2cm} & \hspace{-0.2cm} $\max$(\textsc{glred}$/l$, \textsc{spmv})  \hspace{-0.2cm} & \hspace{-0.2cm} $\max$($4l+1$, $7$) \hspace{-0.2cm} \\
\hline
\end{tabular}
\label{tab:pipelcg}
\end{table}

\subsubsection{Floating point operations per iteration} 

The Conjugate Gradients variants listed in Table \ref{tab:pipelcg} compute a single \textsc{spmv} in each iteration. However, as indicated by the \emph{Time} column, time per iteration may be reduced significantly by overlapping the global reduction phase with the computation of one or multiple \textsc{spmv}s. Time required by the local \textsc{axpy} and \textsc{dot-pr} computations is neglected, since these operations are assumed to scale perfectly as a function of the number of workers.
Alg.\,\ref{algo:PIPELCG} requires $2(l+1)$ \textsc{axpy}s per iteration to update the auxiliary vectors $z^{(0)}_{i-l+2}, \ldots, z^{(l)}_{i}$ using the recurrence relations \eqref{eq:all_z_rec2}-\eqref{eq:specific_z_rec}. The algorithm also computes $(l+1)$ local dot products to form the $G_j$ matrix and uses two \textsc{axpy} operations to update the search direction $p_{i-l}$ and the iterate $x_{i-l}$. In summary, as indicated by the \emph{Flops} column in Table \ref{tab:pipelcg}, the p($l$)-CG algorithm uses a total of $(3l+5)$ \textsc{axpy}s and dot-products, or $(6l+10)N$ flops per iteration (where $N$ is the matrix size).

\subsubsection{Storage requirements}

Table \ref{tab:pipelcg} summarizes the storage requirements for different variants of the CG algorithm. We refer to \cite{ghysels2014hiding} for storage details on the CG and p-CG algorithms. Alg.\,\ref{algo:PIPELCG} stores the three most recently updated vectors in each of the $l+1$ auxiliary bases $Z^{(0)}_j, \ldots, Z^{(l)}_j$ (which includes the bases $V_j$ and $Z_j$). Furthermore, note that the $l$ most recent vectors in the $Z^{(l)}_j = Z_j$ basis need to be stored for dot product computations (see Alg.\,\ref{algo:PIPELCG}, line 23), and that the vector $p_{i-l}$ is also stored (Alg.\,\ref{algo:PIPELCG}, line 30). The p($l$)-CG algorithm thus keeps a total of $4l+1$ vectors in memory at any time during execution (not counting $x_{i-l}$ and $b$). Note that for $l=1$ a minimum of $7$ vectors needs to be stored.

\section{Strong scaling results} \label{sec:strong}

Performance measurements 
result from a PETSc \cite{petsc-web-page} implementation contributed by the authors of the 
p($l$)-CG algorithm on a NERSC distributed memory machine using MPI for communication. 
The latest version of the algorithm, Alg.\,\ref{algo:PIPELCG}, will be included in the PETSc distribution 
under \texttt{> KSP > IMPLS > CG > PIPELCG} from PETSc version 3.11 onward.

\subsection{Hardware and software specifications} \label{sec:hardware}

Parallel performance experiments 
are performed on up to 1024 compute nodes
of the US DoE NERSC ``Cori'' cluster on 16 Intel 2.3 GHz Haswell cores per node. 
Nodes are interconnected by a Cray Aries high speed ``dragonfly'' network. 
More details on the hardware can be found at \url{https://www.nersc.gov/users/computational-systems/cori}
under \texttt{> Cori Configuration > Haswell Nodes}.

Performance measurements 
result from a PETSc \cite{petsc-web-page} implementation of the preconditioned
p($l$)-CG algorithm, Alg.\,\ref{algo:PIPELCG}. 
We use PETSc version 3.10.4. The MPI library used for this 
experiment is Cray-MPICH version 7.5.5. 
The timing results reported are always the most favorable results 
(in the sense of smallest overall run-time) 
over 5 individual runs of each method.
Experiments also show results for classic CG and Ghysels' p-CG method \cite{ghysels2014hiding} as a reference. The p-CG method is similar to 
p($1$)-CG in operational cost and communication overlap (see Table \ref{tab:pipelcg}), although 
it may be numerically less stable than p($l$)-CG, see \cite{cools2018analyzing} for details.

\begin{figure*}[t]
  \centering
  \includegraphics[width=0.48\linewidth]{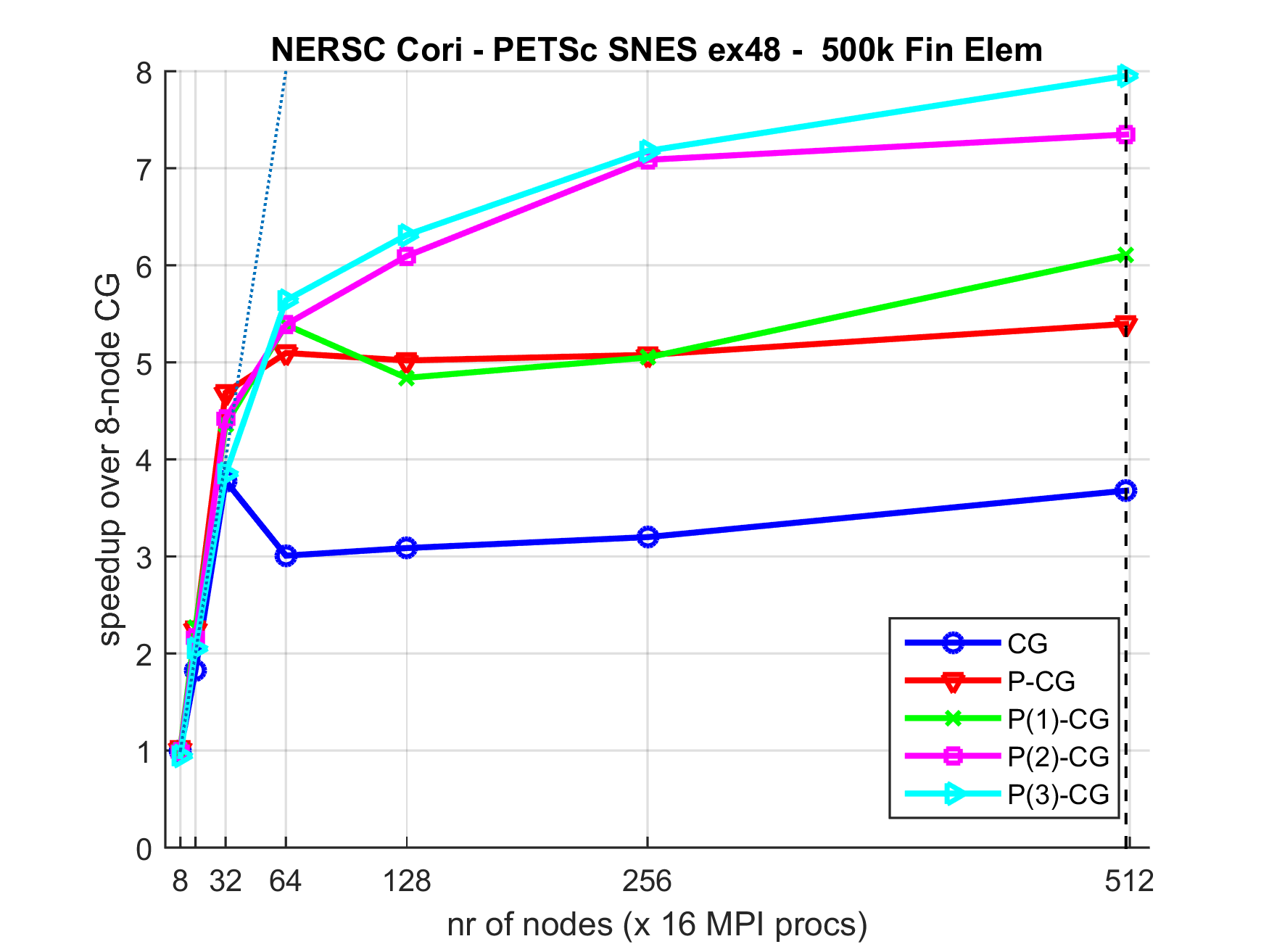} \hfill \includegraphics[width=0.48\linewidth]{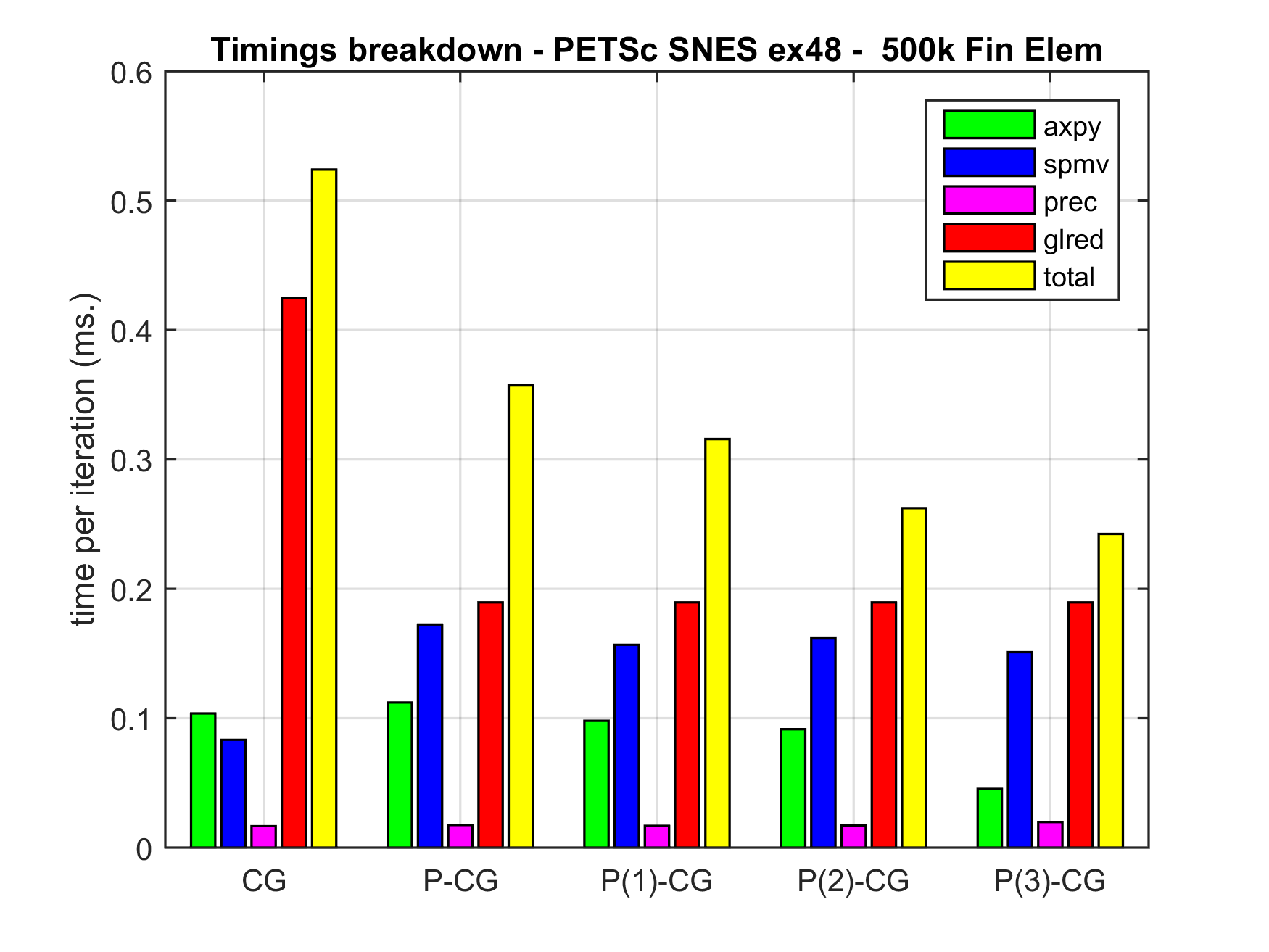} \\
	\includegraphics[width=0.48\linewidth]{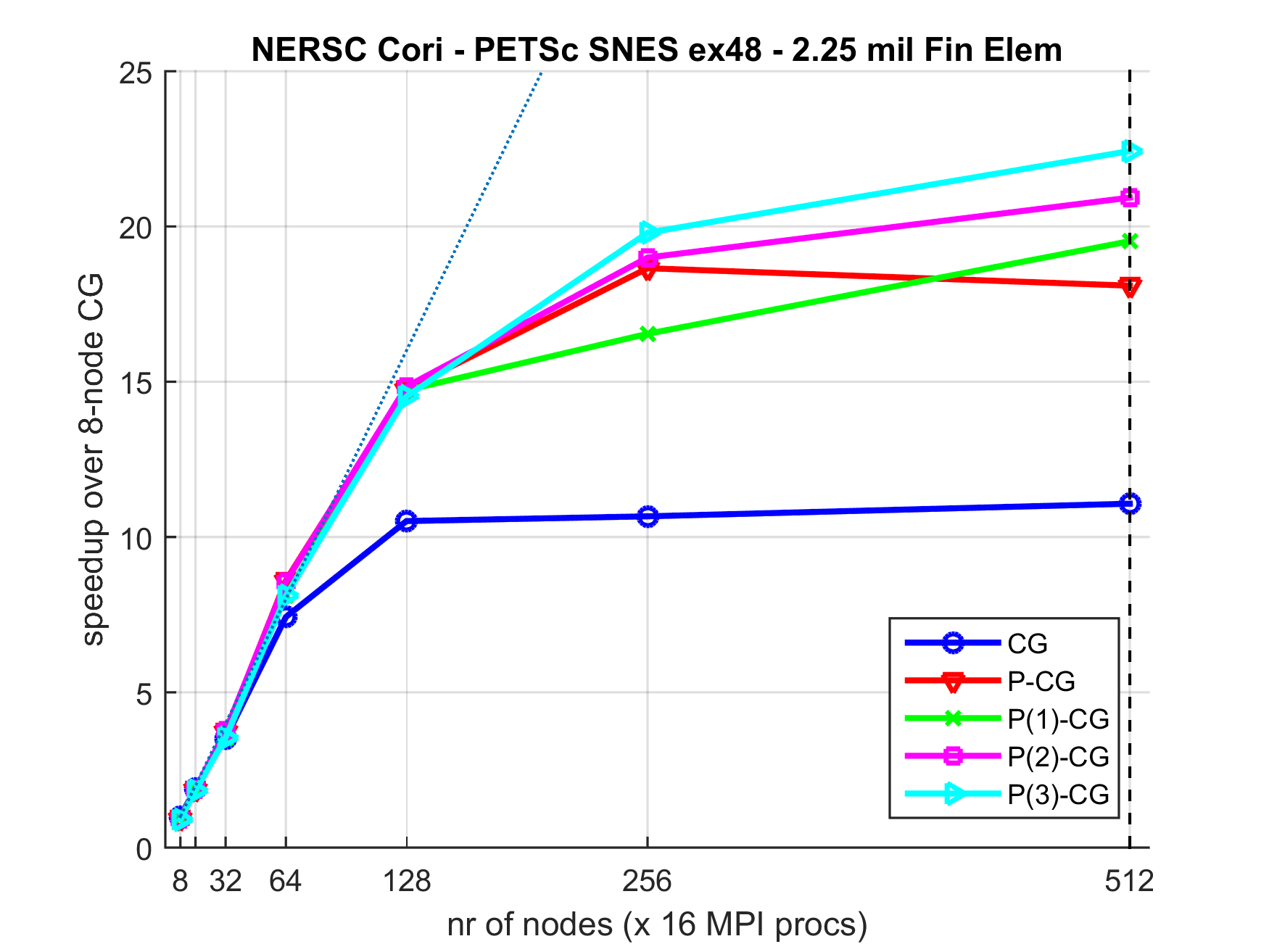} \hfill \includegraphics[width=0.48\linewidth]{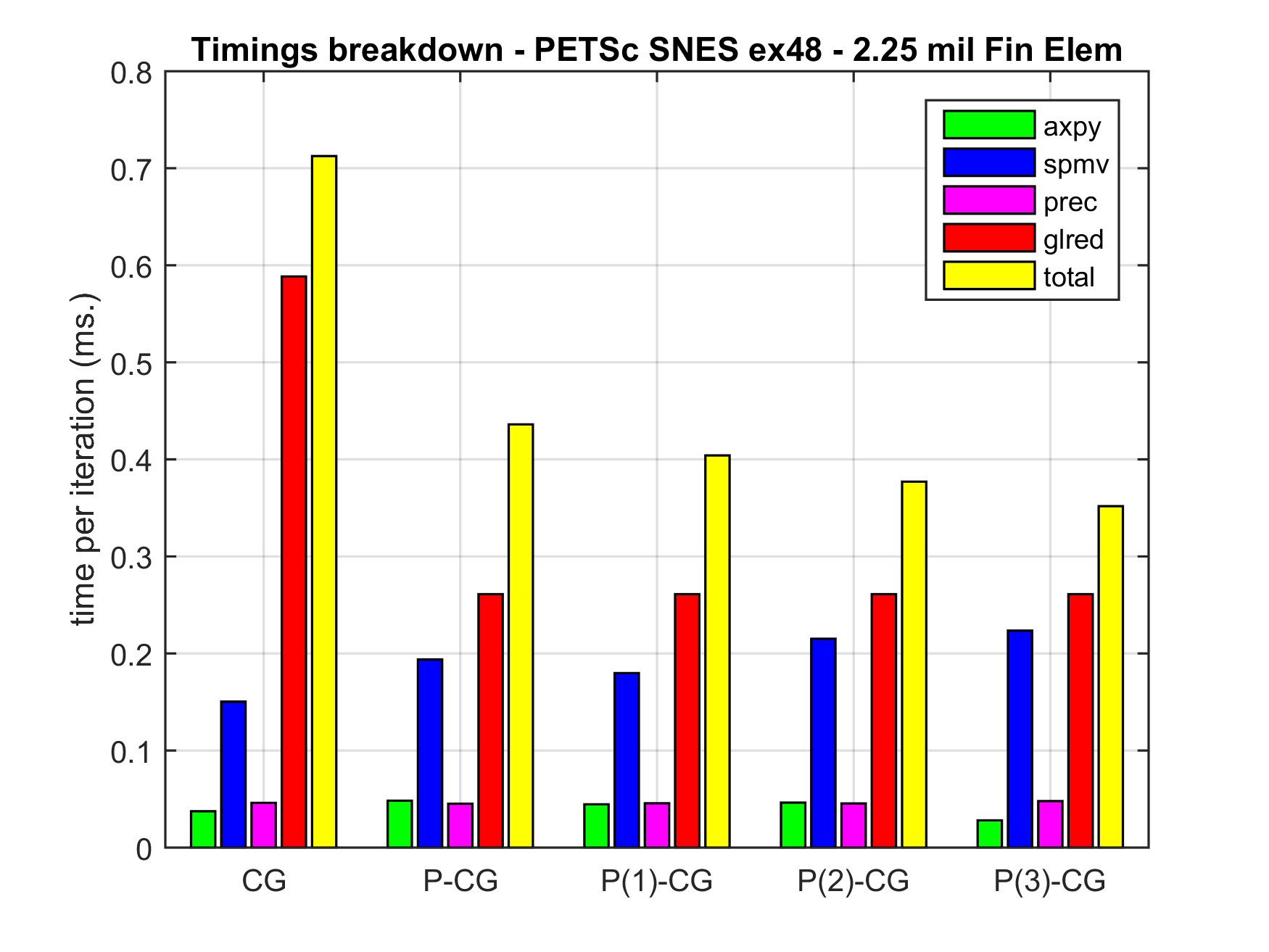} \\
	\includegraphics[width=0.48\linewidth]{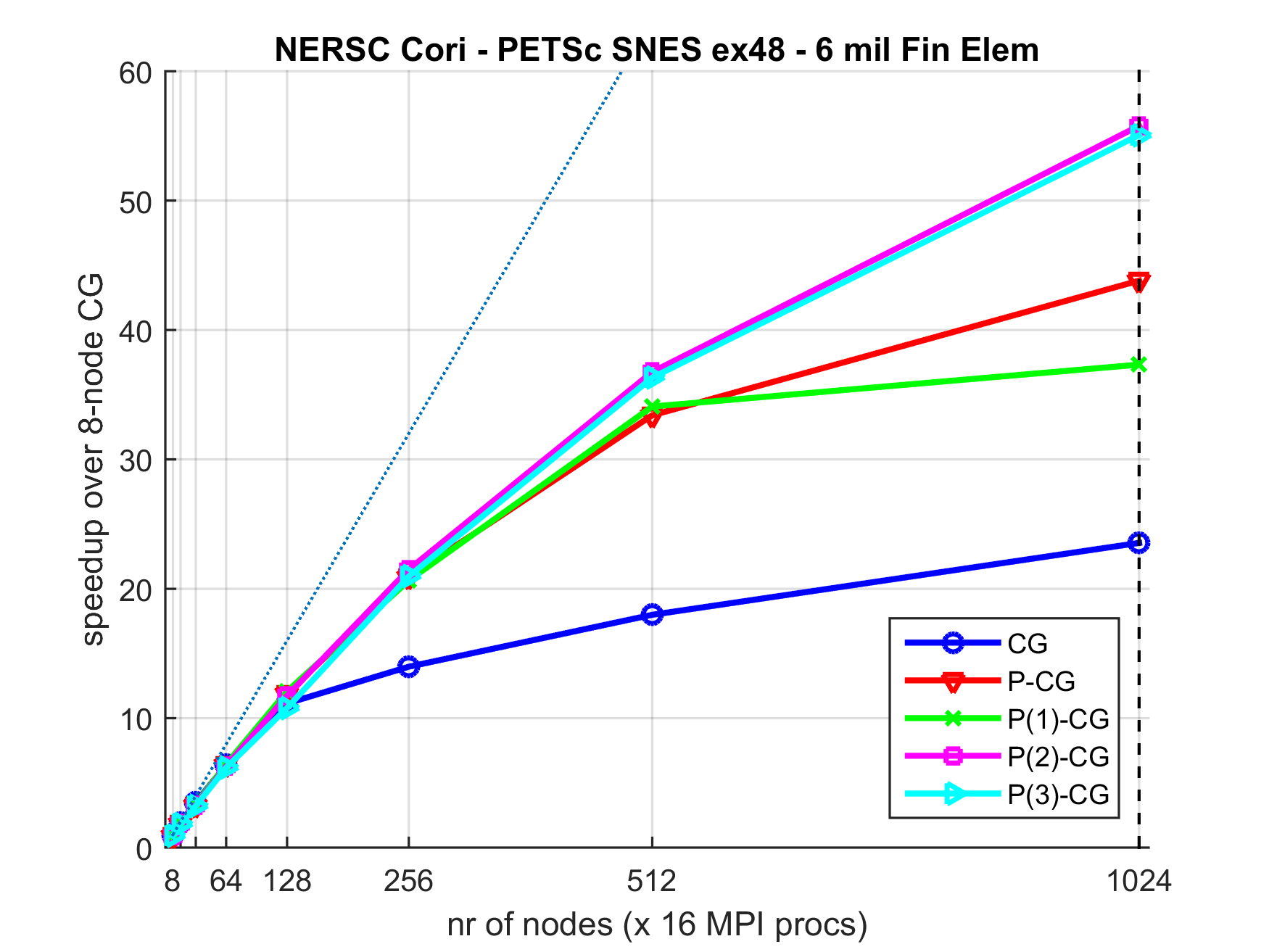} \hfill \includegraphics[width=0.48\linewidth]{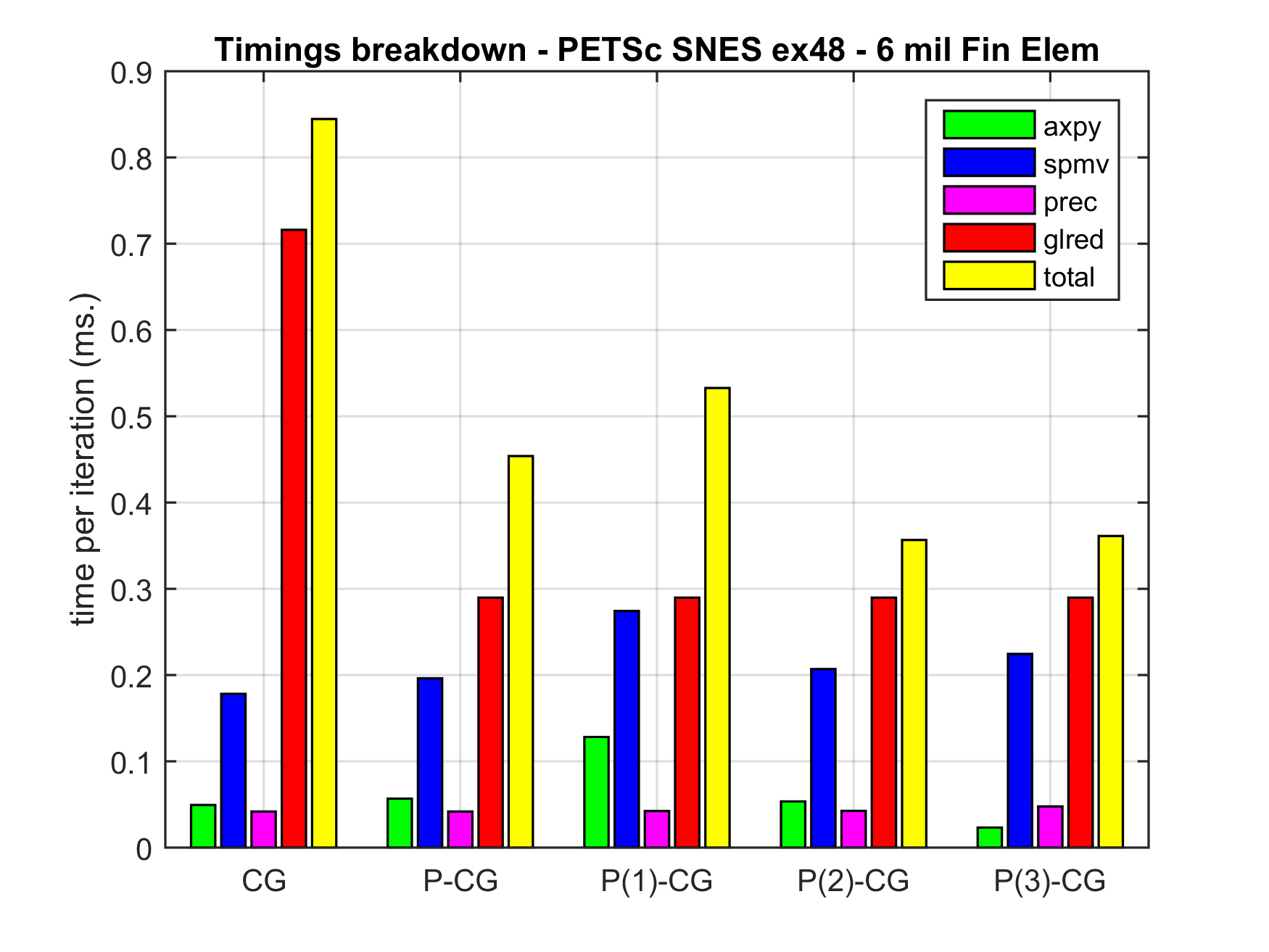} 
  \caption{Non-linear 3D hydrostatic (Blatter/Pattyn) ice sheet flow equations (PETSc SNES ex48).
	Left: speedup over $8$-node classic CG for various pipeline lengths and problem sizes.
	Right: timing breakdown showing the maximum time spent in each kernel. Total timings correspond to the dashed line in the left graph(s).
	Legend: `CG': classic CG, `PCG': Ghysels' pipelined CG, `PCG\,l': l-length pipelined CG (l=1,2,3).}
  \label{fig:strong}
\end{figure*}

\begin{figure*}[t]
  \centering
  \includegraphics[width=0.48\linewidth]{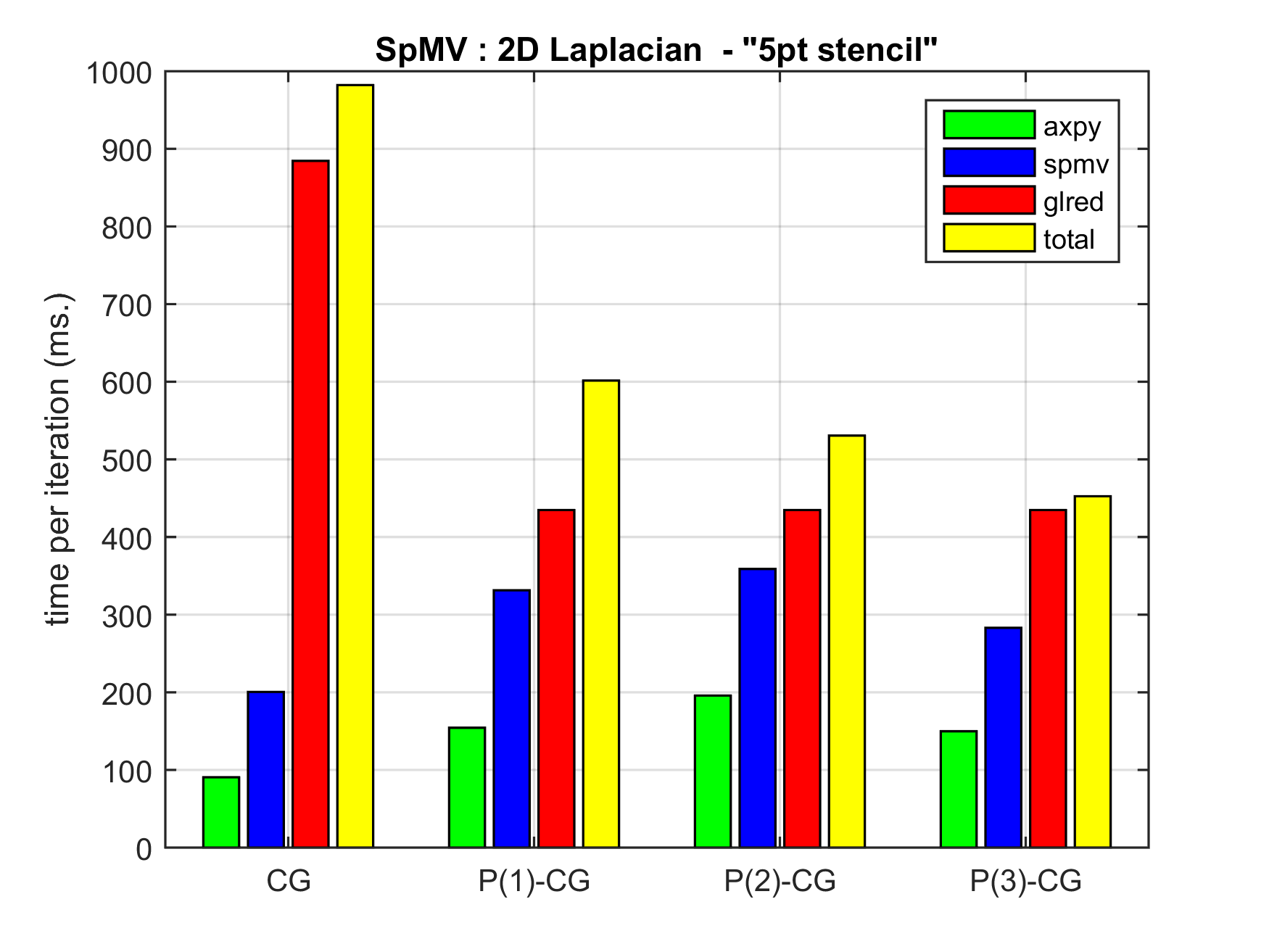} \hfill \includegraphics[width=0.48\linewidth]{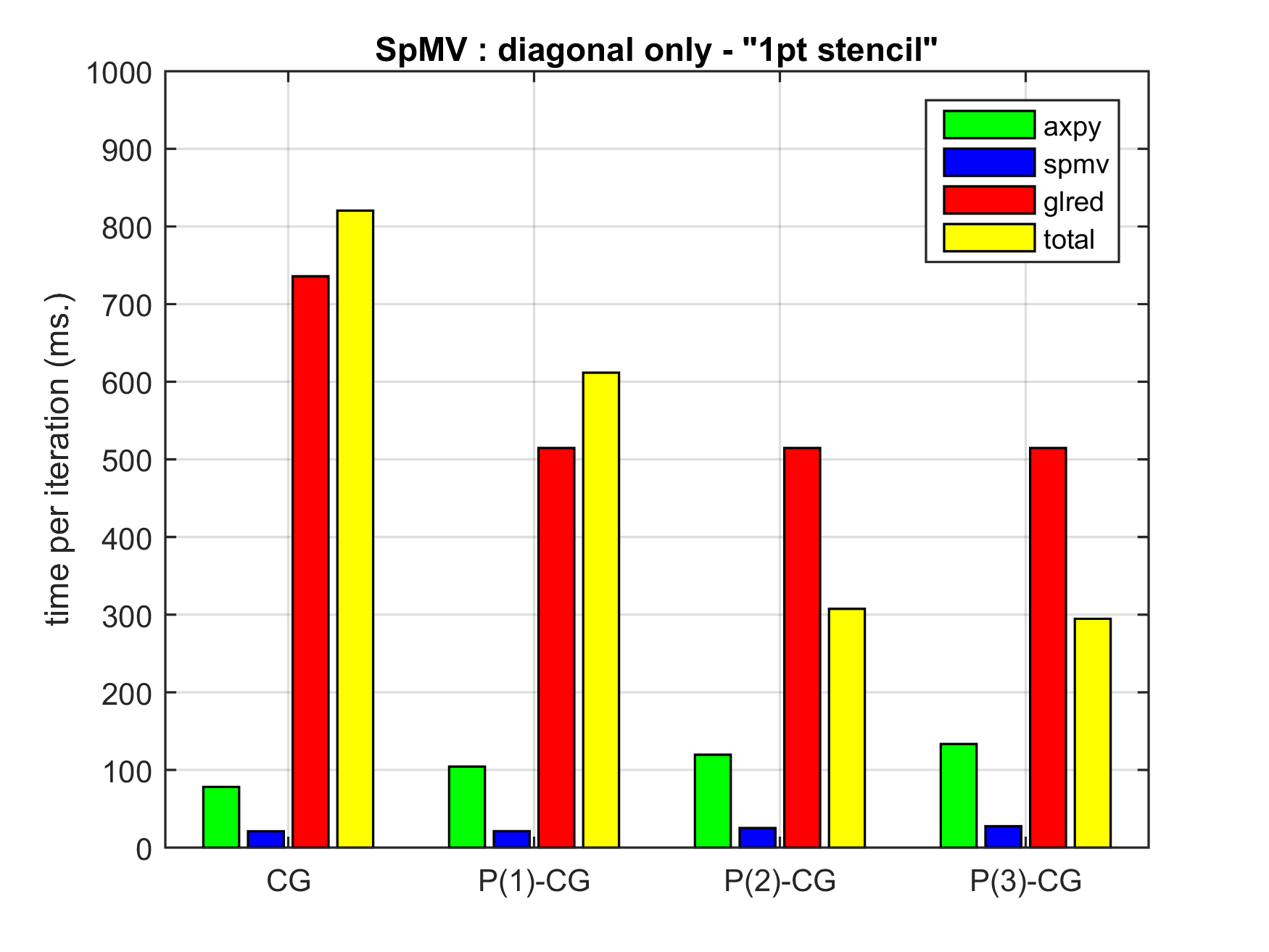}  
  \caption{Timing breakdowns showing the maximum time spent in each kernel for classic CG and p($l$)-CG ($l$=1,2,3). Left: linear 2D finite difference 5-point stencil Laplacian system (PETSc KSP ex2) (see also Fig.\,\ref{fig:schematics}, left panel for interpretation). Right: ``communication bound'' toy problem; diagonal matrix system with identical spectrum and right-hand side to the 2D 5-point stencil Laplacian (see also Fig.\,\ref{fig:schematics}, right panel).}
  \label{fig:lapl}
\end{figure*}

\begin{figure*}[t]
  \centering
  \includegraphics[width=0.48\linewidth]{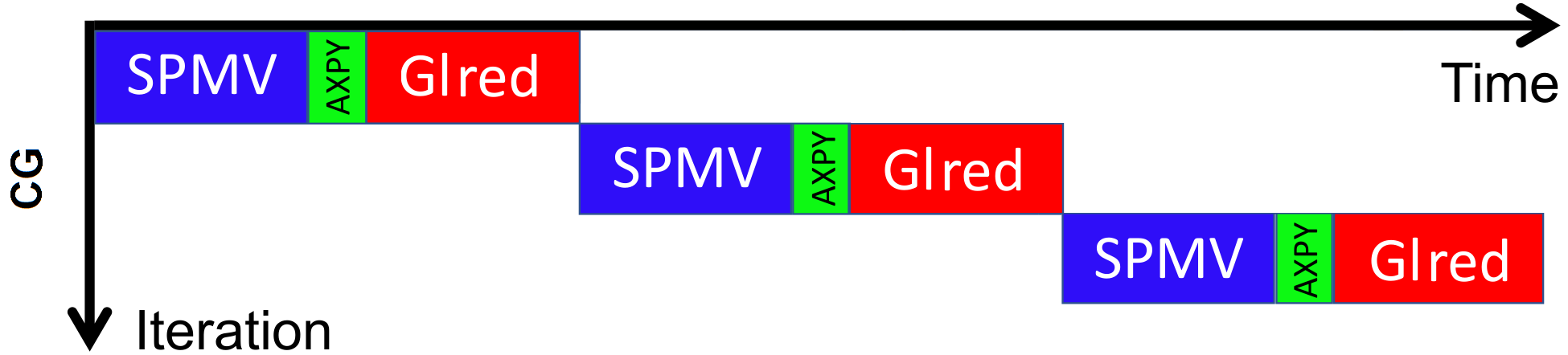} \hfill \includegraphics[width=0.48\linewidth]{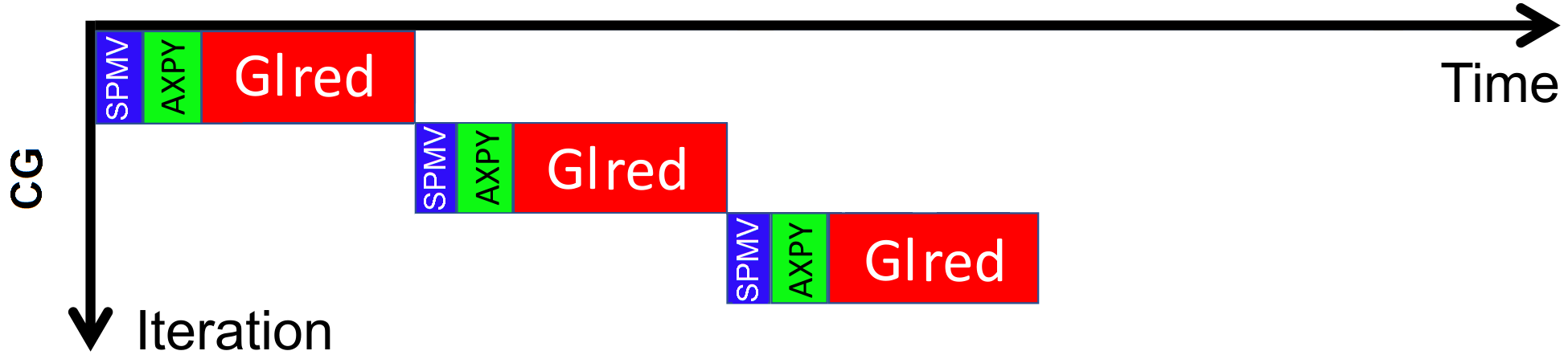}  \vspace{0.1cm} \\
	\includegraphics[width=0.48\linewidth]{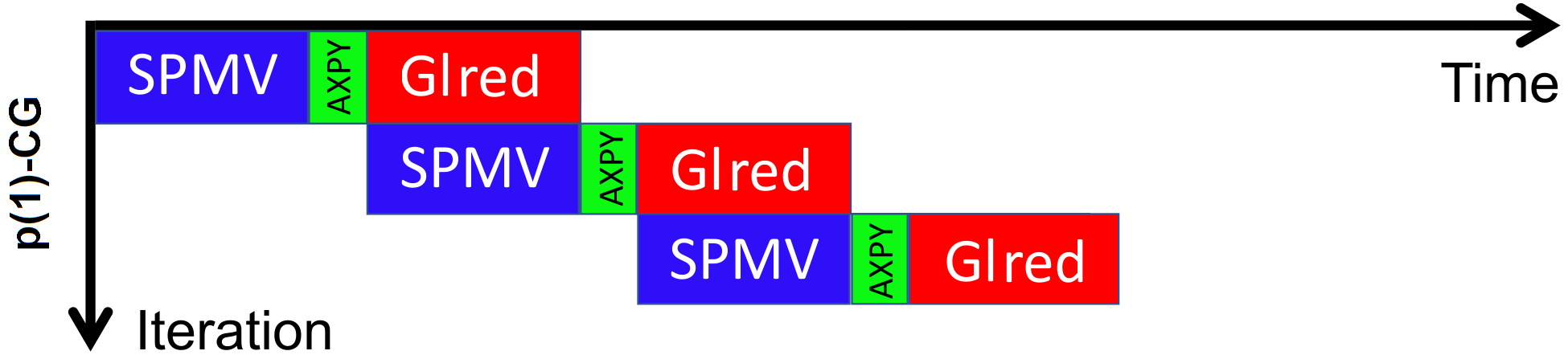} \hfill \includegraphics[width=0.48\linewidth]{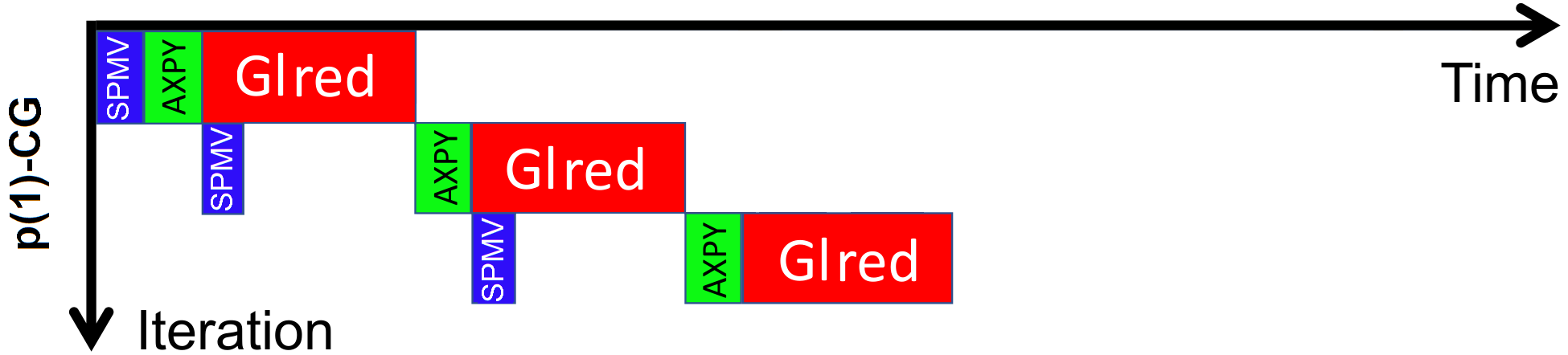}  \vspace{0.1cm} \\
	\includegraphics[width=0.48\linewidth]{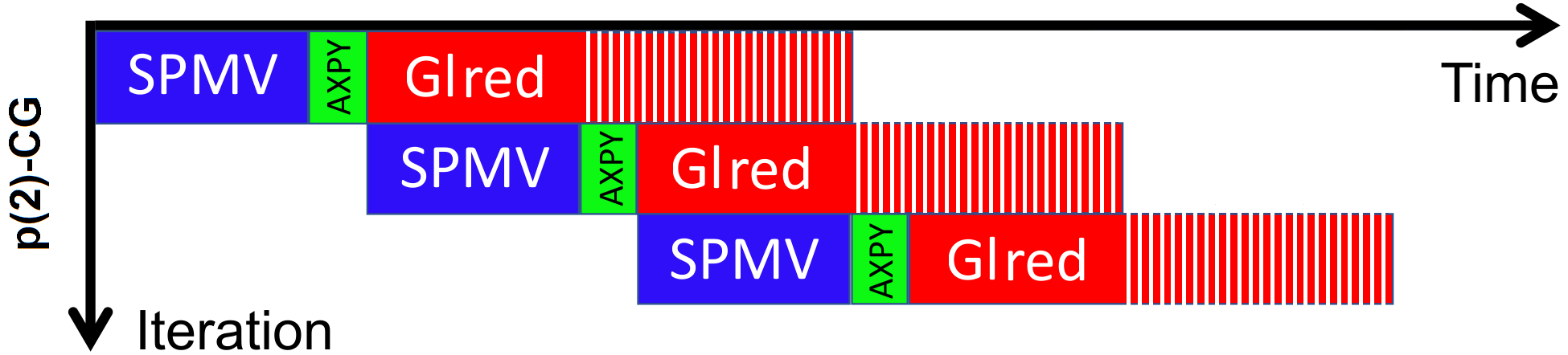} \hfill \includegraphics[width=0.48\linewidth]{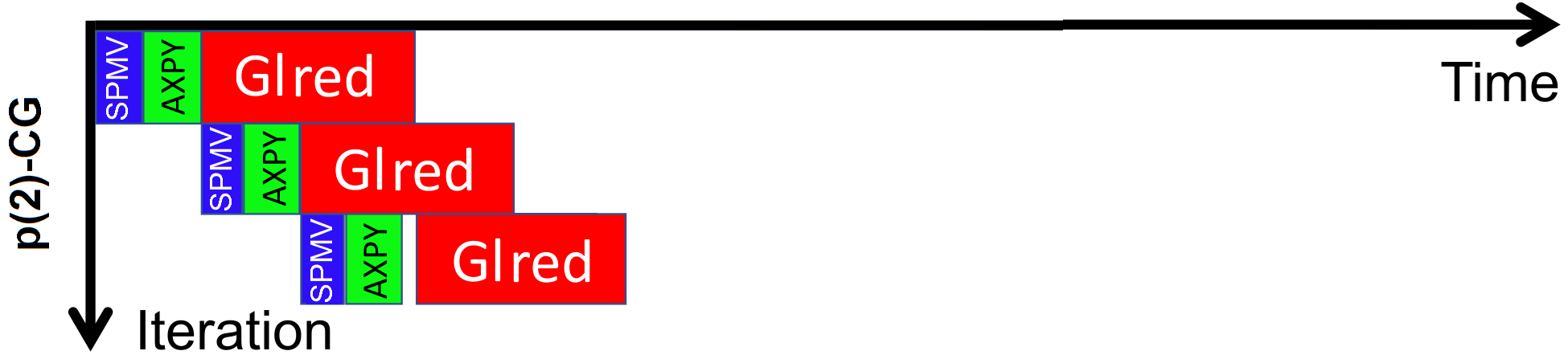} 
  \caption{Schematic representation of the ``communication hiding'' property of the p($l$)-CG method in different scenarios. The work flow for the main \textsc{SPMV}, \textsc{AXPY} and \textsc{GLRED} kernels during three consecutive CG (top), p($1$)-CG (middle) and p($2$)-CG (bottom) iterations are shown. Left: theoretically ``ideal'' scenario for pipeline length $l = 1$, where approximately equal time is spent in \textsc{GLRED} and \textsc{SPMV}. No speedup is expected for pipeline lengths $l \geq 2$ compared to $l = 1$ (see shaded areas). Right: extremely communication-bound scenario where time spent in \textsc{GLRED} is much larger than time of \textsc{SPMV} computation. A significant speedup of p($2$)-CG over p($1$)-CG is expected as a result of the ``staggered'' communication phases.}
  \label{fig:schematics}
\end{figure*}

\subsection{Discussion on experimental data} \label{sec:discussion}

Figure \ref{fig:strong} shows the result of three strong scaling experiments for the 3D Hydrostatic Ice Sheet Flow problem, which is 
available as example $48$ is the PETSc Scalable Nonlinear Equations Solvers (SNES) folder. The Blatter/Pattyn equations 
are discretized using $100\times100\times50$ (top) / $150\times150\times100$ (middle) / $200\times200\times150$ (bottom) finite elements respectively.
A Newton-Krylov outer-inner iteration is used to solve the non-linear problem. The CG methods used as inner solver are combined with a block Jacobi preconditioner (one block Jacobi step per CG iteration; one block per processor; blocks are approximately inverted using ILU). The relative tolerance of the inner solvers is set to $\|b-A\bar{x}_i\|/\|b\| = 1.0$e-$6$. 
For the p($l$)-CG method the Chebyshev shifts $\sigma_0,\ldots\sigma_{l-1}$, see \eqref{eq:chebyshev}, are based on the interval $[\lambda_{\min}, \lambda_{\max}] = [0,2]$.
In summary, the following parameters are provided when calling the p($l$)-CG solver:
\begin{verbatim}
  -ksp_type pipelcg       -pc_type bjacobi 
  -ksp_rtol 1e-06         -ksp_pipelcg_pipel <l> 
  -ksp_pipelcg_lmin 0.0   -ksp_pipelcg_lmax 2.0
\end{verbatim}
The left panel of Figure \ref{fig:strong} shows the speedup of different CG variants relative to the execution time of classic CG on $8$ nodes. The dotted line shows the theoretical optimal speedup based on the $8$-node CG timing. As expected, the classic CG method stops scaling from a certain (problem size dependent) number of nodes onward, i.e.~when the global reduction becomes the dominant time-consuming kernel in the algorithm. The overlap of communication and computations in the pipelined methods successfully ``hides'' the global synchronization latency behind the computational kernels, leading to improved scalability. 

The right panel of Figure \ref{fig:strong} presents a detailed timing breakdown of the individual kernels in the algorithms. The bars represent the maximum timing for a particular kernel over all iterations and over all MPI processes. It is apparent from the red bars that CG features two global reduction phases per iteration whereas the pipelined methods perform only one, see Table \ref{tab:pipelcg} (this feature is sometimes referred to as ``communication avoiding'' in the literature \cite{chronopoulos1989s}). Total times (yellow bars) correspond to the dashed vertical lines in the left panel of Figure \ref{fig:strong}. The total time spent clearly decreases by ``pipelining'' global reductions (but note that the overlap is not explicitly shown in the right panel of Figure \ref{fig:strong}).

Somewhat surprisingly, in most cases the use of longer pipelines ($l > 1$) seems to improve scalability 
compared to the $l = 1$ case. This is unexpected, as Figure \ref{fig:strong} (right) indicates that for these model problems the time spent in global communication is of the same order of magnitude as the time required to compute an \textsc{spmv} (+ \textsc{prec}). Hence, the theoretical model in Table \ref{tab:pipelcg} suggests that a pipeline length $l = 1$ would suffice for a perfect overlap, hiding the entire global reduction phase behind computational work, and that longer pipelines do not necessarily yield any benefit towards performance in this case. Figure \ref{fig:schematics} (left) presents this scenario schematically.

The explanation for the observation that speedup can be achieved for longer pipelines in this particular case is two-fold. 
First, as indicated earlier in this work, a slight benefit may be noticeable from overlapping the global communication with \textsc{axpy}s. For $l = 1$ only the local \textsc{axpy}s on lines 26-33 in Alg.\,\ref{algo:PIPELCG} are overlapped with the \textsc{glred} phase. However, longer pipelines $l \geq 2$ also overlap the \textsc{axpy} operations on line 19-21 of the next $l-1$ iterations with the global reduction phase. This overlap with a (slightly) larger volume of computations may induce a minor speedup when the time spent in the \textsc{axpy} operations is not completely negligible (which is often the case in practice due to read-writes to local memory), although the effective performance gain achieved by this overlap is generally assumed to be quite small.

Second, the overlap of one global reduction by (other) global communication phases may also lead to improved performance, even in the absence of overlap with a significant amount of computational work from the \textsc{spmv}. 
This ``global communication staggering'' effect is assumed to be most apparent in extremely communication-bound regimes, in which the time for a global reduction takes significantly longer than the computation of \textsc{spmv}s and/or \textsc{axpy}s. To validate this intuitively sensible premise, Figure \ref{fig:lapl} presents a simple small-scale experiment for solving a 4 million unknowns system with a finite differences 2D Laplacian five-point stencil (left, PETSc KSP ex2) and a trivial diagonal system (i.e.~a ``one-point stencil'') with the eigenvalues of the 2D Laplacian on the main diagonal (right). Both problems are solved on 128 nodes using classic CG and p($l$)-CG, and are equally challenging from a spectral perspective. 
For the Laplacian (left panel) the benefit of using p($1$)-CG over CG is reflected in the total timings (yellow bars). However, the performance gained from using longer pipelines ($l \geq 2$) is quite limited, since the timing breakdown indicates that \textsc{glred} takes around the same order of time as an \textsc{spmv}. This scenario is represented schematically in Fig.\,\ref{fig:schematics} (left), from which it is clear that pipeline lengths $l \geq 2$ are not useful here since there is no more global communication to overlap. A different view is presented by the diagonal matrix (right panel), which can be interpreted as a toy model for extremely communication-bound scenarios. In this case a \emph{significant} speedup is observed by comparing p($2$)-CG to p($1$)-CG. Longer pipelines with $l \geq 3$ do not improve performance further however. An explanation can again be found in Fig.\,\ref{fig:schematics} (right panel). A small gain in performance can be achieved by using p($1$)-CG instead of CG, which is due to the avoidance of global communication and the overlap of global communication with the \textsc{spmv}. However, contrary to the left-hand side panel, a \emph{significant} performance gain of p($2$)-CG over p($1$)-CG is observed due to the fact that p($l$)-CG with $l \geq 2$ allows to start a new global reduction \emph{before} the global reduction from (the) previous iteration(s) ha(s)(ve) finished.  This overlap or ``staggering'' of the global communication phases for pipeline lengths $l \geq 2$ also allows for more robustness with respect to \textsc{glred} run-time variance. Together with the aforementioned overlap with \textsc{axpy} operations for $l \geq 2$, this ``communication staggering'' explains the speedup that was observed in the strong scaling experiments in Fig.\,\ref{fig:strong} when using p($l$)-CG with pipeline lenghts $l \geq 2$.

\section{Conclusions} \label{sec:conclusions}

The primary aim of this work is to present pipelined Krylov subspace methods as a valuable application of MPI-based programming that is of great practical interest for improving scalability of linear solver codes. The performance gains and strong scaling benefits of the pipelining approach are demonstrated for the deep pipelined CG method on various large scale benchmark problems in this work. In case of optimal overlap of communications and computations, a speedup factor of $\mathcal{O}(l)$ can be observed when comparing p($l$)-CG to classic CG. Similar results can be obtained using MPI-based implementations of other communication-avoiding and -hiding methods, many of which are available in the open source PETSc framework. We believe that pipelined Krylov subspace methods provide an excellent use case for the asynchronous non-blocking communications implemented in the MPI3 standard. We encourage both MPI developers and users from applications to share their insights to further improve MPI functionality for these and other interesting future-oriented applications. Moreover, we encourage MPI developers to address the overhead of thread safety which currently seems to be quite significant as observed in our experiments (but not reported in the current manuscript).

\begin{acks}
The authors are grateful for the funding that supported this work. Specifically,
J.\,Cornelis acknowledges funding by the University of Antwerp Research Council under the University Research Fund (BOF).
S.\,Cools receives funding from the Flemish Research Foundation (FWO Flanders) under grant 12H4617N. 
This work was also supported in part by the U.S. Department of Energy, Office of Science, Office of Advanced Scientific Computing Research, ``Scientific Discovery through Advanced Computing'' (SciDAC) program through the FASTMath Institute under Contract No.~DE-AC02-05CH11231 at Lawrence Berkeley National Laboratory.
\end{acks}

\bibliographystyle{ACM-Reference-Format}
\bibliography{refs2}

\end{document}